# Deep Reinforcement Learning based Model-free On-line Dynamic Multi-Microgrid Formation to Enhance Resilience

Jin Zhao, *Member, IEEE*, Fangxing Li, *Fellow, IEEE*,
Srijib Mukherjee, *Senior Member, IEEE*, Christopher Sticht, *Senior Member, IEEE*

*Abstract*—Multi-microgrid formation (MMGF) is a promising solution to enhance power system resilience. This paper proposes a new deep reinforcement learning (RL) based model-free on-line dynamic multi-MG formation (MMGF) scheme. The dynamic MMGF problem is formulated as a Markov decision process, and a complete deep RL framework is specially designed for the topology-transformable micro-grids. In order to reduce the large action space caused by flexible switch operations, a topology transformation method is proposed and an action-decoupling Q-value is applied. Then, a CNN based multi-buffer double deep Q-network (CM-DDQN) is developed to further improve the learning ability of original DQN method. The proposed RL method provides real-time computing to support on-line dynamic MMGF scheme, and the scheme handles a long-term resilience enhancement problem using adaptive on-line MMGF to defend changeable conditions. The effectiveness of the proposed method is validated using a 7-bus system and the IEEE 123-bus system. The results show strong learning ability, timely response for varying system conditions and convincing resilience enhancement.

## I. INTRODUCTION

HIGH-IMPACT and low-probability events, such as extreme weather events, are occurring with increasing intensity. The extensive damage and subsequent outages of a power system caused by extreme events indicates the necessity of enhancing power system resilience [1]. Microgrids (MGs), which improve the flexibility of power system operation with both grid-connecting and islanding modes, are promising solutions for power grids to withstand unplanned catastrophic events [2]. With widely penetrated distributed energy resources (DERs), advanced metering, communication, and automatic control infrastructures, the distribution system can be easily transformed into self-supported MGs [3]. These self-supported MGs largely benefit power system resilience by improving the restoration capability of the distribution networks [4]-[6] as well as the surviving of critical loads [7].

Since the extreme condition handling ability of MGs has been proven both by academic index [8] and by practical cases (e.g., islanded MGs were successfully survived in Hurricane Sandy [9]), resilience-oriented MG formation, resources allocation, and system operation are widely discussed. The optimal MGs formation strategies were proposed in [4] and [10] to divide the original distribution system (DS) into resilient MGs after major faults of the main grid. Allocable distributed generations (DGs) [11] and remotely controlled switches (RCSs) [12] are highlighted to provide a good planning study of resilient MGs. A transformative architecture for the normal operation and self-healing of multi-MGs was proposed in [13] to improve the system self-healing capability, and [14] used a scheduling-horizon-based optimization scheme to reduce load shedding with reasonable operation cost. In addition, some helpful resilient control strategy was studied to benefit the operation of islanded MGs [15].

To fully utilize DERs to enhance power grid resilience, forming multiple MGs using DGs becomes a promising solution to handle extreme conditions [3]. The essence of multi-MGs formation (MMGF) problem is to identify the desired topology subject to various constraints. For the topology determination problem, mathematical programming [3], [4], [10] and heuristic search approaches [16], [17] are widely used methods. A mix-integer non-linear programming (MINLP) model was built in [3] to sectionalize the outage area into the networked MGs. The MMGF problem was formulated as a mix-integer linear programming (MILP) in [4], and the model is further improved in [10] by reducing both binary and continuous variables. Based on the graph theory, [16] developed a graph-theoretic search algorithm to identify a post-outage distribution system topology. Another heuristic approach was proposed in [17] to approximately solve the MG formation problem of large-scale systems with tractable computation. A good summary of existing methods was provided in [18] especially for the radial topology consideration when MGs are being formed.

The aforementioned MG formation strategies are mainly based on observable system conditions and environments with short-term considerations, while conditions under nature disasters might be uncertain and changeable [7]. The uncertain output of RES-DGs and unexpected damage of grids reduce the efficiency or even damage the initially formed MGs. Therefore, an adaptive and dynamical MG formation strategy is needed to further enhance the resiliency under unexpected system conditions. By continuously interacting with the environment and obtaining feedbacks, the deep reinforcement learning (DRL) method [19] is promising to help the MMGF scheme obtain the adaptability to changeable conditions.

As an efficient solution to handle Markov decision process-

This work was supported in part by National Science Foundation (NSF) award ECCS-1809458 and in part by Oak Ridge National Laboratory (ORNL) (*Corresponding author:* F. Li).

J. Zhao and F. Li are with the Department of EECS, The University of Tennessee, Knoxville. TN 37996, USA.

S. Mukherjee and C. Sticht are with Electrification and Energy Infrastructures Division, Oak Ridge National Laboratory, Oak Ridge, TN 37830, USA.

es (MDPs), DRL methods have become an attractive method for intractable problems in power system. Ref. [20] casted the volt-VAR optimization to a deep Q-network (DQN) framework and finally realized adaptive voltage control under time-varying operating conditions. To achieve real-time service restoration, [21] proposed an imitation learning (IL) framework to improve the training efficiency of DRL methods. In terms of MGs, the DRL method showed satisfying performance in energy management problems [22]-[24]. However, because of the difficulty of ensuring feasible radial topology, few studies discussed the MMGF problem using DRL methods. For the MMGF problem, the action space of DRL methods has exponential growth with the increase of the number of switches, which deteriorates the learning ability of DRLs. Therefore, the DRL based dynamic multi-MG formation is a valuable but challenging problem to be discussed.

For the purpose of realizing on-line dynamic MMGF, a new deep RL based model-free real-time adaptive scheme is proposed in this paper to enhance the resilience in a long time-horizon. First, the dynamic MMGF problem is formulated as a MDP, and the deep Q-learning based RL method is introduced as a promising solution. Second, holding the features of spinning forest, a topology transformation method and an action-decoupling method based on convolutional neural network (CNN) are developed reduce the action space and mitigate tricky topology issues. Finally, several techniques, such as double DQN (DDQN), Epsilon-greedy based exploration and specially designed multi-buffers, are implemented to improve the learning ability of the proposed DRL method.

The contributions of this paper can be summarized as follows: 1) A new DRL supported on-line dynamic MMGF scheme is proposed. A long time-horizon is considered to fully utilize the available DGs under major faults of the main grids. The original problem is reformulated using MDP and a complete DRL framework is specially designed for the topology-transformable MGs. 2) The problem of large action space when applying DRL methods is mitigated. The topology transformation method and the CNN-based action-decoupling Q-network are developed to efficiently handle the issue of exponentially increasing action number. 3) The learning ability of DDQN method is further improved to be the CNN-based multi-buffer double DQN (CM-DDQN) method. The CM-DDQN has strong learning ability and satisfactory computational performance to provide real-time adaptive MMGF strategy according to the newly updated system information.

The rest of the paper is organized as follows: Section II reformulates the dynamic MMGF problem as a DRL based MDP. The topology transformation and the CNN based action-decoupling Q-network methods are provided in Section III. Section IV shows the detailed designs for the training and on-line application of the CM-DDQN based dynamic MMGF scheme. Section V provides case study results and discussions, followed by the conclusions of this work.

## II. MMGF Problem Formation Using DRL Framework

This section introduces the dynamic MMGF problem with a DRL based MDP form. First, the dynamic MMGF problem is formulated to fit into a MDP from. Then, the solution is designed using the deep Q-learning structure with characteristics of the MMGF problem.

### A. Formulate dynamic MMGF as a MDP

The goal of the MMGF is transforming a distribution system (DS) into several self-supported islanded MGs [4]. Under the changeable environment such as extreme weather events, the dynamic MMGF maintains load supply in a time period by adaptively adjusting topologies of multi-MGs. It is a sequential decision-making problem in a multi-step process. At each step, a topology configuration is determined to form islanded MGs by system reconfiguration and splitting based on the current state and the MMGF action of the last step.

Therefore, the dynamic MMGF problem can be described by a MDP which consists of four essential elements: state $S$, action $A$, state transition probability $P$, and reward $r$. In the MDP of the MMGF problem, the agent can be the distribution system operator (DSO). As shown in Fig. 1, the agent takes an action $A_t$ based on the environment's state $S_t$ at each time step $t$. Consequently, the agent gains a reward $r(S_t, A_t)$ and the state transitions to $S_{t+1}$ according to the state transition probability $P(S_{t+1}|S_t, A_t)$. This state–action–next-state process is an interaction between the environment and agent, and it continues until the terminal state or the last step of setup [19].

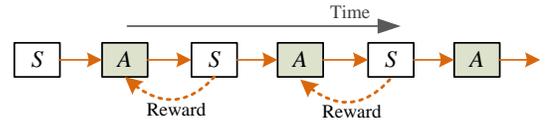

Fig. 1 MDP for the MMGF problem

Assume that the original DS totally has $n$ nodes, $l$ lines, $w$ RCSs ($w \le l$), $n_g$ DG nodes and $n_L$ load nodes and the time horizon is $T$. The binary variable $\alpha$ denotes the close ($\alpha = 1$) or open ($\alpha = 0$) statues of RCSs. Each MG comes from the original DS should be energized by a DG [4], [10].

1) **State**. The state is a part or all of characteristics of the current environment observed by the agent. The state is composed of the current network topology configuration $\boldsymbol{\alpha} = [\alpha_1, ..., \alpha_w]$, active and reactive DG output condition $\boldsymbol{p}_{DG} = [p_{DG,1}, ...p_{DG,ng}]$ and $\boldsymbol{q}_{DG} = [q_{DG,1}, ...q_{DG,ng}]$, load amount $\boldsymbol{p}_L = [p_{L,1}, ...p_{L,ng}]$ and $\boldsymbol{q}_L = [q_{L,1}, ...q_{L,ng}]$ and the time $t$. Accordingly, the state at time $t$ is defined as $S_t = [\boldsymbol{\alpha}_t, \boldsymbol{p}_{DG,t}, \boldsymbol{p}_{DG,t}, \boldsymbol{p}_{L,t}, \boldsymbol{q}_{L,t}]$.

2) **Action**. The action is the reaction of the agent to the current state. In the dynamic MMGF problem, the action at time $t$ $A_t$ can be represented by changing the configuration of networks in $S_t$. Therefore, the action space contains all the combinations of topology configuration $\boldsymbol{\alpha}_t$. However, the action space is quite large with $2^{\wedge w}$ combinations. Since the large space brings troubles of non-convergence of DRL methods, this will be further handled in Section III.

3) **Reward**. The reward is the feedback of the environment after the agent takes some action in a state. For the dynamic MMGF process, the action should first maintain the radial network of each individual MG. Further, the se-curity constraints such as voltage and branch flow limits should be considered. The reward function (1) defines the MMGF problem.

$$r_t\left(S_t, A_t\right) = \begin{bmatrix} f_{topo}\left(\boldsymbol{\alpha}_t\right) - f_{AC}\left(\boldsymbol{p}_t, \boldsymbol{q}_t, \boldsymbol{\alpha}_t\right) \\ -f_{swi}\left(\boldsymbol{\alpha}_0, ..., \boldsymbol{\alpha}_t\right) - f_{pb}\left(\boldsymbol{p}_t, \boldsymbol{\alpha}_t\right) \end{bmatrix} \quad (1)$$

The reward function contains the first term as the reward and the last two terms as penalties of the current action. Specifically, the first term $f_{topo}$ (.) means the reward if switch action $\boldsymbol{\alpha}_{t+1}$ successfully forms multi-MGs with radial networks. The second term function $f_{AC}$ (.) represents the penalty based on AC power flow results. The third function $f_{swi}$ (.) punishes frequent close/open actions of each switch. The last function $f_{pb}$ (.) represents the penalty for insufficient power supply.

Note that the reward is an immediate feedback for one-step. However, in the dynamic MMGF problem, the long-term feedback with cumulative reward is concerned. Thus, the return (2) is defined which is the accumulation of the current reward and the discounted future rewards.

$$R_t = \sum\nolimits_{k=0}^{T-1} \gamma^k r_{t+k} \quad \gamma^k \in \left[0, \ 1\right] \quad (2)$$

where $\gamma^k$ is the discount factor.

### B. Dynamic MMGF using deep Q-learning

Different from classic dynamic programming methods, the DRL method does not require explicit policies and value functions for MDPs as well as the complete knowledge of MDPs. [19]. Therefore, it becomes a promising approach to solve the complex dynamic MMGF problem. The well-trained DRL method can quickly provide on-line scheme for dynamic MMGF, which helps the original grid give an adaptive reaction under changeable environments. This further enhances the system resilience.

As introduced in Subsection II-A, the dynamic MMGF problem is with the discrete action space and continuous state space. This feature makes it suitable to apply the deep Q network (DQN) method [25]. The DQN method is a combination of deep neutral network (DNN) and Q-learning which updates the action-value function iteratively. For a policy $\pi$, define the action-value (Q-value) function as (3),

$$Q^\pi\left(S_t, A_t\right) = \mathbb{E}\left[R_t \big| S_t, A_t, \pi\right] \quad (3)$$

where Q-value is the expected discounted reward for executing action $A_t$ at state $S_t$ and following policy $\pi$. The objective of Q-learning is to estimate the value for an optimal policy. It has been proven that an optimal policy can be derived from the optimal Q-values $Q^*$ $(S_t, A_t)$ = max $Q^\pi(S_t, A_t)$ by selecting the action with the highest Q-value in each state [26]. Therefore, the agent can decide how to properly perform actions by learning the Q values. For the dynamic MMGF problem, the Q-value can guide proper MMF decisions in the MDP introduced in subsection II A.

Based on the *Bellman equation*, (3) can be further represented as a recursive format (4). As a form of *temporal difference* (TD) learning, Eq. (5) can update the Q-value towards the targeted one with the learning rate $\eta$.

$$Q^\pi\left(S_t, A_t\right) = \mathbb{E}\left[r_t + \gamma Q^\pi\left(S_{t+1}, A_{t+1}\right)\right] \quad (4)$$

$$Q\left(S_t, A_t\right) \leftarrow Q\left(S_t, A_t\right) + \eta \begin{bmatrix} r_t + \gamma \max\limits_{A_{t+1}} Q\left(S_{t+1}, A_{t+1}\right) \\ -Q\left(S_t, A_t\right) \end{bmatrix} \quad (5)$$

Theoretically, the convergence of the iterative process is guaranteed, which means the $Q^*$ $(S_t, A_t)$ can be found [26]. The agent can be guided to optimally perform actions using $Q^*$ $(S_t, A_t)$. However, the Q-value is hard to be functioned in the dynamic MMGF problem, and it is hard to provide a reasonable Q-table because of continuous state space. Therefore, the Q-value function is approximated with deep neural network (DNN) parameterized by $\theta$ (6). As such, the original Q-learning method is transformed into the DQN method [25].

$$Q\left(S_t, A_t\right) \approx Q\left(S_t, A_t \big| \theta\right) \quad (6)$$

The DNN based Q-value is updated with loss function representing the mean-squared TD error, as shown in (7). For the MDP, the first two terms in (7) represent the direct reward of the current action and the potential value of the current action for future MDP, respectively. Together, they measure the value of the current action. The last term directly generates the value of current action using Q-network. By minimizing the loss function, the Q-network gradually learns to generate Q-values guiding proper MMGF schemes.

$$L\left(\theta\right) = \left[r_t + \gamma \max\limits_{A_{t+1}} Q\left(S_{t+1}, A_{t+1} \big| \theta\right) - Q\left(S_t, A_t \big| \theta\right)\right]^2 \quad (7)$$

## III. ACTION GENERATION OF DYNAMIC MMGF PROCESS

In this section, the problem of large action space when applying DRL methods is mitigated. First, a topology transformation method is used to handle the radial topology requirement. Then, the CNN-based action-decoupling Q-value is designed to further handle the large action number.

### A. Search space reduction of spinning forest

Since the action space contains all the combinations of topology configuration $\boldsymbol{\alpha}_t$, the original scale of the action space is $2^{N_W}$. This exponentially increasing action space brings troubles to the convergence of (5), and it contains tremendous infeasible network configurations because the radial network of each MG should be maintained [4], [18]. Moreover, the infeasible topology and the computation burden make power flow calculation-based environment interaction difficult to be performed. Therefore, topological issues need be addressed.

From the point of topology changing, splitting a tree will lead to several trees. Therefore, the MMGF problem includes the reconfiguration and splitting of the original DS. As shown in Fig. 2, all the reasonable radial MGs can be found by: 1) reconfiguring the original radial DS by switching operations, and 2) splitting the reconfigured DS by opening any closed switches.

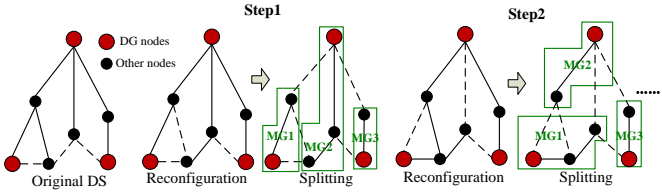

Fig. 2 The process of multi-MGs formation

The essence of the reconfiguration and splitting process is to find all the spanning forests of a network topology. Since it is intractable to directly trace all the spanning forests, a topology transformation method, as shown in Fig. 3, is applied to simplify the problem.

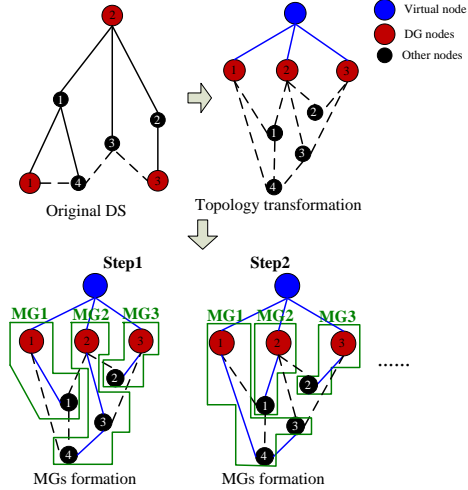

Fig. 3 Topology transformation for MMGF

First, the DG nodes are picked up and connected using a virtual node (the blue node in Fig. 3). Accordingly, the spanning forest problem can be simplified as the spanning tree problem. In this way, finding MMGFs is equivalently transformed into the problem of finding radial networks of the new topology. The radial topology of the new topology can be ensured by two conditions [18]: 1) $n$-$n_g$ RCSs switch on (the virtual node is not included in $n$) and 2) all the nodes must be connected.

The first condition reduces the action space from the exponential form ($2^w$) to the polynomial form $C_{n-n_g}^w$. This is how it works: the original DS has $n$ nodes, $n_g$ DG nodes and $w$ RCSs in total. So, the number of original possible combinations is $2^w$. After applying the proposed topology transformation method, the new topology has $n + 1$ nodes (including the virtual node) and $w + n_g$ lines (including the virtual lines between the virtual node and DG nodes). To ensure radial topology, the new network should have $n$ lines in operation. Because $n_g$ virtual lines have already been in operation, $n$ - $n_g$ RCSs should be switched on. Therefore, the problem is transformed into picking up $n$ - $n_g$ RCSs in $w$ RCSs, which has $C_{n-n_g}^w$ possible combinations. This is significantly less than the original $2^w$ combinations.

Based on the first condition, the second one can be used to check the feasibility of the network.

## B. CNN-based action-decoupling Q-value

The key point of the DQN method is to learn Q values by building and training the neural network. Regarding power related applications, the DQN method is normally based on the structures of ANN [27] or DNN [18], [28]. However, as defined in (3), the Q-value is the expected accumulation of the discounted reward functions. The reward function for the MMGF problem heavily depends on the AC power flow calculation which has sparse function relationship [29], [30], and the neighboring branches have stronger topology relationship. Therefore, the neutral network is organized using CNN which has strong automatic feature learning ability in processing data with a grid-like topology with sparsely connected features. The CNN based neural network is promising to better learn the Q values of the MMGF problem.

The data preprocessing for the deep CNN of the Q-value is based on the new topology in Fig. 3. The input data contains node active and reactive power injection vectors $\boldsymbol{P^0}$ and $\boldsymbol{Q^0}$, branch resistance and reactance elements $\boldsymbol{R^0}$ and $\boldsymbol{X^0}$ elements, switch open/close statues $\boldsymbol{W}$ and frequencies of switching operation $\boldsymbol{F}$. In order to ensure the consistent dimension, $\boldsymbol{P^0}$ and $\boldsymbol{Q^0}$ are extended with $l$-$n$ zeros (if $l > n$) to become $\boldsymbol{P}$ and $\boldsymbol{Q}$, and $\boldsymbol{R}$ and $\boldsymbol{X}$ are formed by considering the original branch parameters $\boldsymbol{R^0}$ and $\boldsymbol{X^0}$, as well as the switch status $\boldsymbol{W}$. The input is organized as $[\boldsymbol{P}; \boldsymbol{Q}; \boldsymbol{R}; \boldsymbol{X}; \boldsymbol{F}]$.

Fig. 4 shows the process of generating Q-value using the CNN structure. Assuming the first convolutional layer is with filters of the size [3, 3, 1, 12] where the first three numbers are the height, width and depth of one filter and the last one is the number of filters. The zero-padding is applied to maintain the original size of the input data. The filter of $Conv2$ has the size [3, 3, 12, 24]. Hence, the output of $Conv2$ has the size [5, $w$, 24] and it further flattened as a vector with the size [1, 5×$w$×24] and goes through a $FC1$ layer. Using a matrix of the weight parameters with the size [5×$w$×24, $w$] and a [1, $w$] parameter, the output will become a vector with the size [1, $w$].

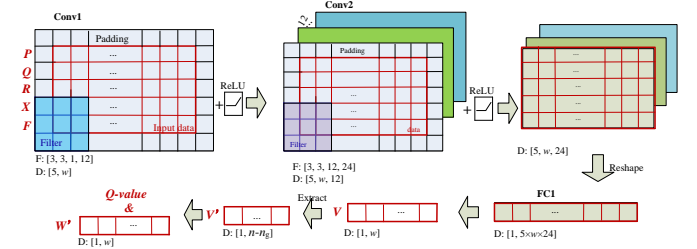

Fig. 4 Structure of CNN for Q-value

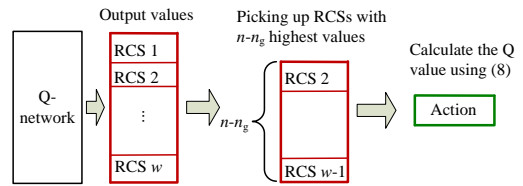

Fig. 5 Design of action-decoupling Q-value

Although the action number has been reduced by the to-pology transformation method, the number of output data of CNN still be large if each action is considered as an output. Therefore, instead of taking each output data as an action, an action-decoupling method is designed, as shown in Fig. 5, by setting the CNN output $V$ as values of switches. Accordingly, the Q-value does not take a specific value of the DQN output; instead, it is set to the average value of a selected subset in the CNN output data $V$. That is, top $n$-$n_g$ values in $V$ are selected and extracted as $V$ and the Q-value (8) is the average of them. In general, the action-decoupling method takes the outputs $V$ of CNN as values for each switch, selects the closing switches, and then calculates the Q-value. When transforming $V$ to $V'$, we may record the switch index numbers of the top $n$-$n_g$ values, set 1 (closed) to these switches and 0 (open) to the remaining switches. Then, the new switch statues is $W$ obtained by closing the selected switches and opening the rest ones.

$$Q\left(S_{t+1}, A_{t+1} \middle| \theta\right) = \frac{\text{sum}\left(V'\right)}{n - n_g} \quad (8)$$

A $min\_max\_scaler$ transformation is applied to normalize the input data $P$; $Q$; $R$; $X$ and output data $V$. The frequencies of switching operation in $F$ are normalized by dividing the total step number. Switch open/close statues in $W$ are originally binary. Through the normalization, the values of the data are within the range [0, 1] which helps create a more regular search region for faster convergence of the algorithm. The loss function used to train the CNN is given by (7). The reward function for the MMGF problem is formulated considering the topology requirements, power balance, voltage, branch flow and switch operation times which will be further functioned in Section IV-B.

## IV. LEARNING AND APPLICATION OF CM-DDQN

The learning and application of the proposed CM-DQN method are discussed in this section. First, techniques for better learning are incorporated in the DRL method. Then, detailed designs of the reward function and the RL process for dynamic MMGF are presented. Finally, the entire method as well as the on-line application framework are concluded.

### A. Techniques for better learning

The key to realizing the DRL based dynamic MMGF is to let the Q-network learn the proper reactions in the MDP. Many techniques have been studied for the efficient learning of DQN. The experience replay, Epsilon-greedy based exploration and fixed network are three most efficient ones.

#### 1) Experience replay and multiple buffers

The DQN learns the Q-value based on the experience. However, for the MDP, the previous experiences are overwritten with new experiences. This largely reduced the data efficiency. Therefore, the experience replay method [25] is applied to memorize the experiences and re-train the Q-network. Accordingly, each experience can be used repeatedly, and bias due to correlation between training samples can be eliminated.

The experience replay consists of a *memory* part and a *replay* part. The *memory* contains a list of previous experiences and observations to re-train the Q-network. In the MDP, state $S_t$, action $A_t$, reward $r_t$, next state $S_{t+1}$ and topology surviving condition $D$ are appended to the *memory*1. As long as the memory stores enough experiences, the replay part is activated.

$$memory1 = \left[..., \left(S_t, A_t, r_t, S_{t+1}, D_t\right), ...\right] \quad (10)$$

Since the learning of the Q-network largely depends on the experience, reserving a part of the good experience helps the Q-network learn proper reactions. For the MMGF problem, good experiences are with high reward values and feasible topology. Accordingly, the multi-buffers are constructed with extra *memories* to reserve good experience extracted from the original *memory*. The replay part randomly extracts some experiences from buffers to organize *minibatch* to train the Q-network.

$$memory2 = \left[..., \left(S_t, A_t, r_t^*, S_{t+1}, D_t\right), ...\right] \quad (11)$$

$$memory3 = \left[..., \left(S_t, A_t, r, S_{t+1}, D_t^*\right), ...\right] \quad (12)$$

The training of the Q-network is enhanced using the *minibatch* in the MDP.

$$minibatch = \begin{bmatrix} \text{random.sample}\left(memory1, batch\_size1\right), \\ \text{random.sample}\left(memory2, batch\_size2\right), \\ \text{random.sample}\left(memory3, batch\_size3\right) \end{bmatrix} \quad (13)$$

#### 2) Epsilon-greedy based exploration

Since the CNN based Q-network is initialized with random weights and bias, its performance is hard to be satisfying in early stage. Therefore, instead of selecting actions directly use the not well-trained Q network, it is better to try all possibilities before it starts to see the patterns. The randomly selection of actions is called 'exploration', while the prediction using DQN is 'exploitation'. The *Epsilon* represents the exploration rate which is a certain percentage to randomly select its actions. The Epsilon-greedy method uses an annealing $\varepsilon$ value to guide 'exploration' and 'exploitation'. As shown in (14), with the constant k controlling the annealing speed, the $\varepsilon$ value gradually decreases after DQN training begins. In each MDP step, the agent randomly extracts a value in [0, 1]. Then, the agent selects the action with the largest Q value if the random value is less than $\varepsilon$; otherwise, a random action will be selected.

$$\varepsilon = \begin{cases} k\varepsilon & \varepsilon > \varepsilon_{\min} \\ \varepsilon_{\min} & \varepsilon \le \varepsilon_{\min} \end{cases} \quad (14)$$

#### 3) Fixed Q network (Double DQN)

Because the update process (5) picks up the maximum Q value of the next state, the overestimation becomes a long-standing problem for all Q-learning based algorithms. In order to address this issue, double DQN (DDQN) is proposed [30] with better results on ATARI 2600 games than other Q-learning based methods. Therefore, it is applied in the deep Q learning based MMGF scheme.

The DDQN has two separate neutral networks: the original Q network and target Q (T-Q) network, which decouples the action selection and action evaluation. the original Q network is used to select the action with maximum Q value while T-Q network evaluates the Q value of the selected action. The T-Q network is a fixed network which is not updated in the Q network updating process. The fix feature enhances the efficiency and stability in the learning process. The loss function (7) is adjusted into (15) accordingly.

$$
L_t(\theta) = \begin{cases} \left[ r_t - Q\left(S_t, A_t \mid \theta\right) \right]^2 & (t = \text{T}) \\ \left[ r_t + \gamma \max_{A_{t+1}} \text{T-}Q\left(S_{t+1}, A_{t+1} \mid \theta^{Tar}\right) \right. \\ \left. - Q\left(S_t, A_t \mid \theta\right) \right]^2 & (t \neq \text{T}) \end{cases} \tag{15}
$$

Based on the original deep Q learning structure, the extra designs 1-3) gives the DQN good performance in dealing with the overestimation issue and providing better learning process. Till now, all the designs for the CM-DDQN method are presented.

### B. CM-DDQN learning process

Defining suitable reward function is an indispensable part to complete the learning process of DRL methods. The detailed reward function (1) is shown in (16) to help determine the Q network of the dynamic MMGF problem.

$$
r_t(S_t, A_t) = \begin{cases} \left[ f_{\text{topo}}(\alpha_t) - \sum_{i \in M_g} f_{\text{AC},i}(P_t, Q_t, \alpha_t) \right. \\ \left. -f_{\text{swi}}(\alpha_0, ..., \alpha_t) - f_{\text{pb},t}(P_t, \alpha_t) \right] & S_{\text{topo}} = 1 \\ f_{\text{topo}}(\alpha_t) & S_{\text{topo}} = 0 \end{cases} \tag{16}
$$

where $S_{\text{topo}}$ is the signal to show whether the switch action $\alpha_t$ successfully forms multi-MGs with radial networks, and $M_g$ is the set of newly formed MGs. As shown in (17), if $S_{\text{topo}} = 1$, $f_{\text{topo}}(.)$ provides the reward w; otherwise, $f_{\text{topo}}(.)$ gives punishment $-w$ and the 'game over' signal of an MDP is triggered because of the infeasible topology. Functions related to AC power flow $f_{\text{AC},i}(.)$, switch status $f_{\text{swi}}(.)$ and power balance $f_{\text{pb}}(.)$ are further explained in (18), (19) and (20), respectively.

$$
f_{\text{topo}}(\alpha_t) = \begin{cases} w & S_{\text{topo}} = 1 \\ -w & S_{\text{topo}} = 0 \end{cases} \tag{17}
$$

$$
f_{\text{AC},i}(P_t, Q_t, \alpha_t) = \sum_{j \in i} p_{\text{vol},j} + p_{\text{loss},i} + \sum_{l \in i} p_{\text{bran},l} \tag{18}
$$

$$
f_{\text{swi}}(\alpha_0, ..., \alpha_t) = \sum_{j \in w} p_j \tag{19}
$$

$$
f_{\text{pb},i}(P_t, \alpha_t) = \begin{cases} 0 & M \leq 0 \\ M & M > 0 \end{cases} \tag{20}
$$

$$
M = \sum_{j \in i} p_{\text{L},j} + p_{\text{loss},i} - \sum_{j \in i} p_{\text{DG},i} \tag{21}
$$

The AC power flow related function $f_{\text{AC},i}(.)$ provides the punish value of forming MG $i$. It contains penalties of voltage violation $p_{\text{vol},j}$ (22), system power loss of $p_{\text{loss},i}$, and branch overflow $p_{\text{bran},l}$ (23). The $f_{\text{swi}}(.)$ function punishes frequent close/open actions of each switch using (24) which works if switch $j$ exceeds the allowed number of operations in the whole dynamic MMGF process. As shown in (20) and (21), $f_{\text{pb}}(.)$ gives punishes to MG $i$ if it has power deficiency.

$$
p_{\text{vol},j} = \begin{cases} 0 & 0.95 \leq V_j \leq 1.05 \\ \text{p}_{vol} & otherwise \end{cases} \tag{22}
$$

$$
p_{\text{bran},l} = \begin{cases} \left(L_l - L_{\text{upp},l}\right) / \text{B}_{\text{base}} & L_l > L_{\text{upp},l} \\ 0 & L_l \leq L_{\text{upp},l} \end{cases} \tag{23}
$$

$$
p_j = \begin{cases} 0 & \sum_{s=1}^{t} \left| \alpha_{j,s} - \alpha_{j,s-1} \right| \leq S_{\text{upp},j} \\ \sum_{s=1}^{t} \left| \alpha_{j,s} - \alpha_{j,s-1} \right| / \text{N}_{\text{step}} & otherwise \end{cases} \tag{24}
$$

where $V_j$ is the voltage amplitude of node j, $L_l$ and $L_{\text{upp},l}$ are respectively the absolute value and upper bound of branch power of line $l$, $\text{B}_{\text{base}}$ is the base value to standardize the branch flow penalty, $S_{\text{upp},j}$ is the allowed operation number of switch $j$, and $\text{N}_{\text{step}}$ is the required step of the dynamic MMGF process.

---

**Algorithm:** CM-DDQN learning process

**Input:** DG generation data set $\{ \boldsymbol{p}_{\text{DG}}^1, ..., \boldsymbol{p}_{\text{DG}}^{\text{T}} \}$ and $\{ \boldsymbol{q}_{\text{DG}}^1, ..., \boldsymbol{q}_{\text{DG}}^{\text{T}} \}$, load data set $\{ \boldsymbol{p}_{\text{L}}^1, ..., \boldsymbol{p}_{\text{L}}^{\text{T}} \}$ and $\{ \boldsymbol{q}_{\text{L}}^1, ..., \boldsymbol{q}_{\text{L}}^{\text{T}} \}$. Initial topology of the original DS $\boldsymbol{\alpha}^0 = [\alpha_1^0, ..., \alpha_w^0]$. Apply topology transformation in Fig. 3.

**Output:** well-trained action-value Q network

**S1**: Initialization. Initialize Q network and T-Q network with same random weights and bias. Initial replay *memory*1-3 with capacity *maxlen*. Set $D^{\text{step}} = 0$. Set batch size, Episode $M$, step number T and Epsilon-greedy parameters.

**S2: for** Episode from 0 to M **do**
    Initialize state $S^0 = [P^0; Q^0; R^0; X^0; F^0]$
    **for** Step from 1 to T **do**
        Perform Epsilon-greedy, and randomly select an action $\alpha^{\text{step}}$ or $\alpha^{\text{step}} = \text{argmax} [Q(S_t, A_t)]$.
        Calculate reward value (16). If topology infeasible, set $D^{\text{step}} = 1$.
        Organize new state $S^{\text{step+1}}$. Note that $\boldsymbol{R}$, $\boldsymbol{X}$ and $\boldsymbol{F}$ are updated according to $\alpha^{\text{step+1}}$ while $\boldsymbol{P}$ and $\boldsymbol{Q}$ follow DG generation and load data sets.
        Add record $[S^{\text{step}}, \alpha^{\text{step+1}}, r^{\text{step+1}}, S^{\text{step+1}}, D^{\text{step}},]$ in *memory*. Add record to *memory*2 if $r^{\text{step+1}} \geqslant r_{\text{std}}$. Add record to *memory*3 if $D^{\text{step}} = 0$.
        **If** topology is infeasible/ $D^{\text{step}} == 1$ **do**
            Update T-Q-CNN as Q-CNN
            **Break;**
        **End if**
        **If** conditions for replay is satisfied **do**
            Randomly select batch size records from *memory*. Train Q network (Q-CNN) using loss function (15).
            **If** Step = T **do**
                Update T-Q-CNN as Q-CNN
            **End if**
        **End if**
    **End for**
**End for**
**S3**: Obtain the Q-CNN.

The CM-DDQN learning process for the MMGF problem is provided the above algorithm description. Therein, there are hard constraints and soft constraints. A "game over" is triggered if any hard constraint is violated, while soft constraint violations lead to certain consequences instead of immediate "game over". In the dynamic MMGF problem, the hard constraint is the feasible topology requirement. If the switch actions cannot ensure radial network of each MG, there will be a 'game over' and the Markov process is directly ended. The voltage limit, branch flow limit, switching number limit and power balance limit belong to soft constraints which forms penalties to organize the reward value of the switch on/off decisions in current step.

### C. Deep RL based dynamic model-free MMGF scheme

The whole DRL based on-line dynamic MMGF process is shown in Fig. 6. Since the feasible topology is hard to be quickly learned, a pre-training part is prepared to make the Q-network capture some topology and power flow related features of the original system. This helps the DRL scheme to be directly used under emergencies or major fault conditions of the main grid. In a major fault event process, the pre-trained Q-network will quickly provide the MMGF scheme to take fully use of the current available DERs. Meanwhile, new experience will be recorded in the buffer and the Q-network can be further updated when the training condition is trigged. Specially, in the on-line application process, a 'do-nothing' module [32] can be added to ensure the topology feasibility. The actions produced by the Q-network will be re-checked using the topology check module in the environment. The 'do-nothing' module is triggered if the re-organized network is infeasible, and the produced action will not be implemented to maintain the feasible radial network of the last step.

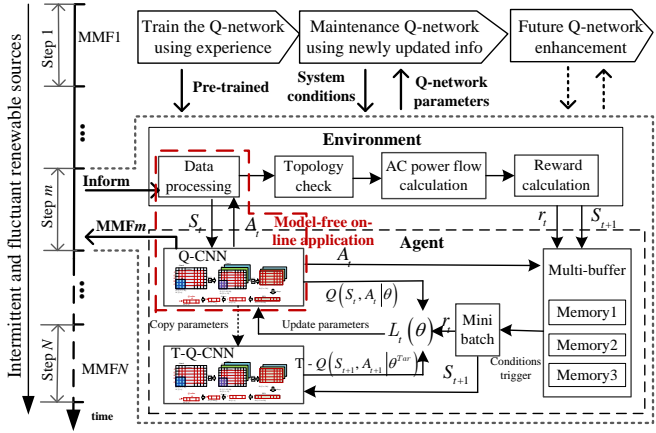

Fig. 6 CM-DDQN based dynamic MMGF process

As such, the DRL based dynamic MMGF scheme can satisfy the real-time computation requirement of the on-line application because the RL agents obtain switch on/off decisions very quickly through simple numerical calculations. Without the actual power system modeling or power flow equations, the on-line application can be performed in a model-free way. On the other hand, the agent can keep on learning new experiences according to newly updated system condition and resilient reactions, which improves the adaptability to handle changeable event conditions. The essence of the proposed dynamic MMGF scheme is to enhance system resilience by flexibly changing topology to form different self-supported MGs according to the newly updated system conditions.

### V. Case Study

In this section, the training and application performances of the proposed CM-DDQN based dynamic MMGF scheme is demonstrated. The dynamic MMGF scheme is compared with the conventional initially-formed MGs schemes [4], [10]. The proposed CM-DDQN is compared with the DDQN [28].

Two systems are used: the 7-bus system with 2 DGs and the IEEE 123-bus system with 12 DGs integrated. The time horizon is set as 200 minutes, and each MG formation transformation step is set as 10 minutes. The switch limit for each remotely-controlled switch (RCS) is set as 4. The initial value of $\varepsilon$ is 1 and $\varepsilon_{min} = 0.1$. Filters of the CNN are with the size [5, 5, 1, 12] and [5, 5, 1, 24]. The uncertain data of DG outputs follow a 3 sigmas normal distribution with 20% forecast error from the expected values. The DRL codes and the corresponding environment are written and compiled in Python 3.7 while the CNN is built using TensorFlow 2.2. and Keras 2.4. Pypower 5.1 is applied to solve the power flow calculation in the environment. All simulation studies were conducted on a computer with Intel® Core (TM) i7-8550U CPU and 16 GB RAM.

### A. DRL based dynamic MMGF process using 7-bus system

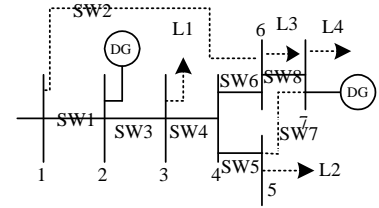

Fig. 7. The 7-bus system

#### 1) Pre-training of deep RL

For the 7-bus system, the episode number for training is 500, and the DG output values for each step of the 500 episodes are randomly generated. The numbers of input and output data are 5×8 and 8. Since the feasible topology is regarded as the hard constraint, the topology condition is the primary concern in the whole process. Note that the switch on/off decisions are obtained from the Q-network output value of each switch. Using the method introduced in Section III-B, the switches with the top 5 highest values according to the CNN output data $V$, are regarded as switch on, and the rest are switch off.

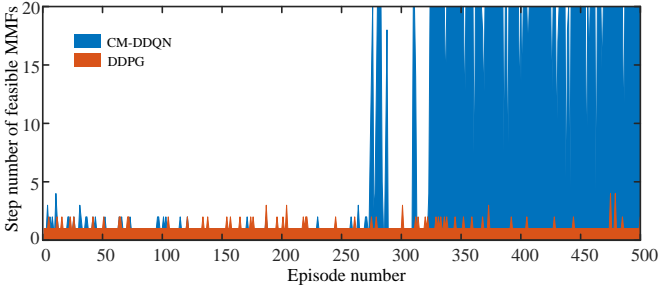

Fig. 8. Successful MMGF steps in the training process

The deep deterministic policy gradient (DDPG) method is applied in this case to compare with the proposed CM-DDQN. The actor-network of DDPG has 8 outputs representing 8 switches, and the output of the actor-network is limited to 0~1. If the output value is larger than 0.5, then, switch on; otherwise, switch off.

The number of steps with radial networks of each MG is shown Fig. 8. A 20-step feasible topology condition means a successful dynamic MMGF process, while any infeasible topologies in the process directly lead to a 'game over' which means the end of an episode. As can be observed, after almost 350 episodes, the proposed DRL method successfully learned how to form feasible topologies by providing switch on/off decisions in the dynamic multi-step process.

After the switch statues are determined, the reward function (16) can be calculated according to the topology check and AC power flow calculation in the environment. The convergence process of the return, which is obtained using the Q value and the reward, is shown in Fig. 9. The values of the return are organized according to ten times of training. The maximum return value and the minimum one are extracted from ten separate trainings and become the upper and lower bounds, which form the light blue area in Fig. 9. As shown in the figure, from 0 to 330, the return first went through an exploration process with low values; then, it increases rapidly with the episode increasing from 330 to 370. After that, the return value becomes relatively stable with small oscillations. Since the return value contains a comprehensive consideration of the topology condition, voltage violation, branch overflow, switch number limits and power balance, the convergence means the Q-network can reasonably judge the performance of an action. As the comparison shows, the output values of DDPG critic-network keeps below zero, which means an unsuccessful learning process.

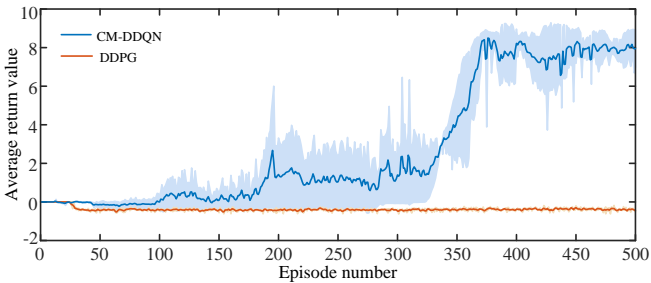

Fig. 9. Convergence process of the return

2) *Comparison of performances of different schemes*

After the Q-network learned to proper judge the performance of actions, a series of uncertain output of DGs in the MDP is randomly extracted to test the performance of the proposed method. The total DG output in 20 steps is shown with the green line in Fig. 10. With the same DG output conditions, the proposed DRL based dynamic MMGF scheme (Dy_MMGF) is compared with the conventional mathematical programming based MMGF scheme (Con_MMGF) [4], [10]. In steps 1-16, both schemes hold all the loads. However, the conventional MMGF scheme (Con_MMGF) sheds loads in steps 17-20. If the dynamic MMGF scheme (Dy_MMGF) still holds all the loads. With the whole dynamic process considered, the DRL based scheme properly dispatches RCS actions, and adjusts the topology based on newly updated system conditions. According, it shows better load-supplying ability.

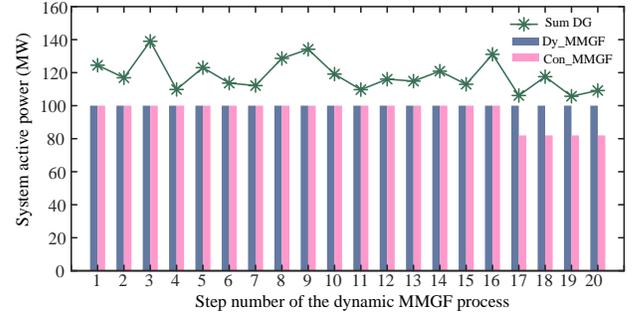

Fig. 10. DG output and reserved load amount in 20 steps

As shown in Fig. 11, the proposed deep RL method changed the formation of MGs, and this action avoided load shedding. The uncertain output of DG1 reduced to 45MW ~ 49MW in steps 17-20. If the original MG formation is reserved, the MG1 will have to shed load L2 to ensure power balance and voltage security. Then, the Con_MMGF scheme has 18 MW load reduction while the proposed Dy_MMGF scheme successfully holds all the load in the entire system.

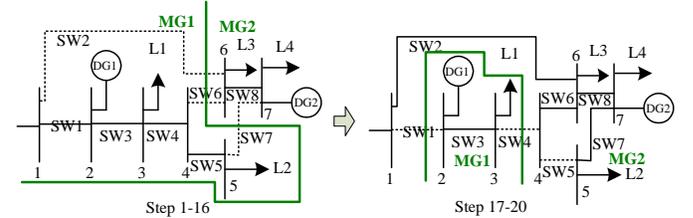

Fig. 11. Dynamic Multi-MG formation

B. *Performance of CM-DQN in IEEE 123-bus system*

The basic data of IEEE 123-bus system is obtained from [33]. The modified IEEE 123-bus system has 124 buses (including the main substation), 11 DGs, 125 lines and 13 RCSs. The original possible topology is 8192, and the number is reduced to 3432 using the topology transformation method in Subsection III-A. The size of input data is 125×5, while the output size is 13×1.

1) *Comparison of learning abilities of DRL methods*

Using the modified IEEE 123-bus system, the learning ability of the proposed CM-DDQN is compared with the DDQN [28] method and CNN based DDQN (C-DDQN) without the multi-buffer part.

As shown in Fig. 12, the large action space brings troubles to the DDQN. Although it has the tendency to learn the proper behavior with a growing successful MMGF step in the early period (about 1-50 episodes), the features are lost in the following training episodes, and finally lead to a failed training process. Without the multi-buffer part, the learning ability of C-DDQN is unstable. Although it has learned proper actions with 20 successful MMGF steps in some episodes (e.g., about 430 and 740 episodes), it quickly lost the features and finally led to an unstable training process. With an improved CNN structure and a multi-buffer design, the proposed CM-DDQN successfully captures the feasible topology feature after about 1000 episode's training.

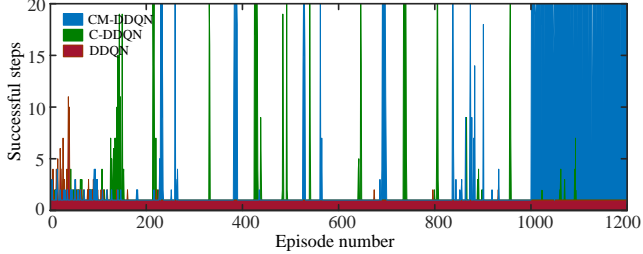

Fig. 12. Comparison of successful MMGF steps of two methods

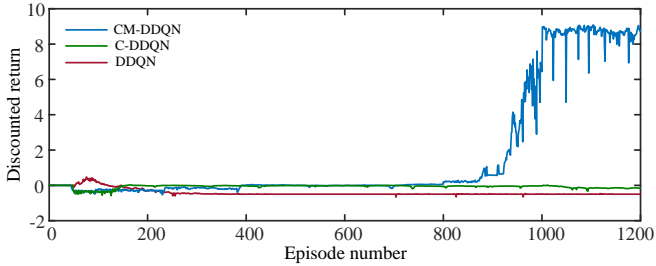

Fig. 13. Convergence process of the return

The corresponding return values of the three methods are demonstrated in Fig. 13. The DDQN and C-DDQN method fails to provide proper evaluation for actions, while the return of the CM-DDQN method reaches to a relatively stable condition. That is because the CNN has strong automatic feature learning ability in processing data with a grid-like topology with a sparse connectivity, while the designed reward value is based on the AC power flow in the sparsely connected power system. In addition, the multi-buffer design provides stable and satisfying experiences for the learning process, which avoids losing the good results explored previously. Note that the large fluctuation in steps 1000-1200 is caused by the exploration design (random action generation) with the lowest 1% probability. Even the Q-network has already well-trained in the training process, the surviving of feasible topology may be reduced by following a randomly generated infeasible action.

### 2) Computation performance of the CM-DDQN

Take a test data with a series of uncertain DG outputs as an example. The proposed CM-DDQN ensures radial networks with three topology forms (topo1, topo2 and topo3) in the 20 steps. Fig. 14 shows the worst voltage condition of the three topology forms. Because the voltage belongs to the soft constraint, there are few violations in the whole process with the proposed CM-DDQN. However, these slight violations are less than 1% of rated bus voltages and are easy to be eliminated by local compensates. The corresponding power losses and most frequently operated switches 8-135 and 13-152 are presented in Fig. 15. The topology transformation happened at step1, step 3 and step 17. The most frequently operated switch has 3 actions in the whole process, which is within the limit of 4 actions.

To verify the on-line application performance of the CM-DDQN method, 100 times of complete MDP (2000 steps) are randomly extracted from the test data. The result is listed in Table I. For the two systems, there are no hard constraint violations, which means that the proposed method successfully ensure the topology feasibility. For the soft constraints, the small-scale systems all have satisfaction performance without any violations. The large-scale system has 4 steps of voltage violations of 5 buses. However, the violations are all within 0.005 p.u., since serious voltage violation brings high penalties for the reward function. In the 100 times of MDP, there are 3 times of switch violations with 5, 5 and 6 times of switch operations, respectively. This phenomenon can be mitigated by increasing the corresponding penalty in the reward function. However, it is not suggested to enhance the consideration of switching actions because it will not lead to security problems and over-focus on soft constraints will influence the performance regarding hard constraints.

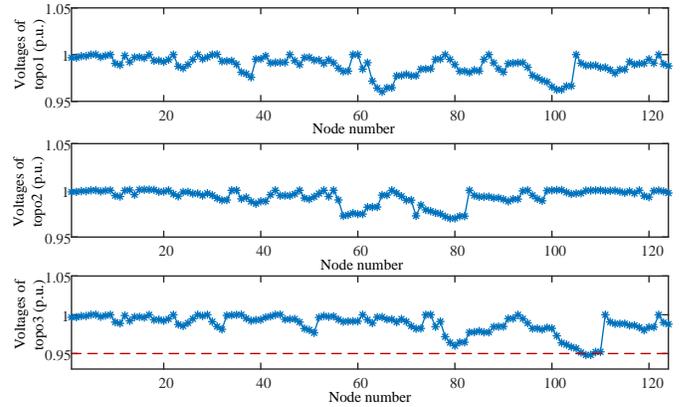

Fig. 14. Voltage conditions of three MMGF topologies

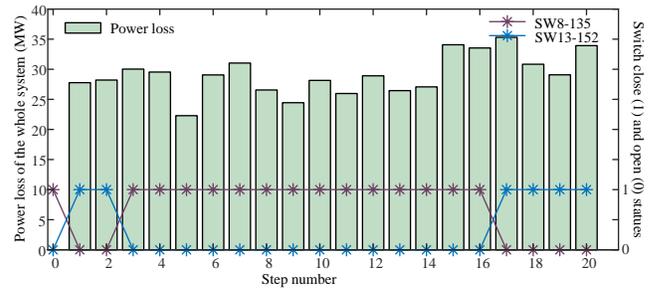

Fig. 15. Power loss and switch conditions of the whole process

In on-line applications, the CM-DDQN provides feasible MMGF strategy with about 0.1 *s* computational time even for the 123-bus system. Although the system scale extended from the 7-bus to 123-bus, the computation time just slightly increases because the proposed method derives results from the model-free DRL structure rather than via the actual power-system model or power flow equation. This feature supports the dynamic on-line MMGF scheme by providing timely topology-adjusting strategy according to newly updated system conditions.

TABLE I. RESULTS OF 100 TIMES OF MDP

| Case | Training time (h) | On-line computation time | Hard cons violation | Voltage violation | Switch violation | Brach flow violation |
|---|---|---|---|---|---|---|
| 7-bus | 5.27h | 0.06s | 0% | 0% | 0% | 0% |
| 123-bus | 58.08h | 0.104 | 0% | 0.2% | 3% | 0% |

## VI. Conclusions

The changeable conditions caused by extreme events reduce the efficiency or even damage the initially-formed MGs. In order to improve the adaptability of MMGF scheme, this paper proposes a new DRL based dynamic on-line MMGF scheme. A DRL based MDP is designed to provide solution for the transformable MMGF in a long-time horizon. A topology transformation as well as a CNN based action-decoupling Q-value is developed to handle large action space problem. The DDQN is improved to formulate the CM-DDQN which enhances the learning ability for large-scale systems. The case study results demonstrate that the proposed dynamic on-line MMGF scheme enhances the resilience by holding all the loads having feasible topology adjustment and the proposed CM-DDQN has strong learning ability, distinguished computation speed in real-time, and satisfactory security guarantee.